\documentclass[aps,10pt,onecolumn,notitlepage,letterpaper,superscriptaddress,nofootinbib,longbibliography,prd]{revtex4-2}
\usepackage[T1]{fontenc}
\usepackage[utf8]{inputenc}
\usepackage{lmodern}
\usepackage[margin=27mm]{geometry}
\usepackage{amsmath,amssymb,mathtools}
\usepackage{mathrsfs}
\usepackage{graphicx}

\usepackage{microtype}
\usepackage{xcolor}
\usepackage[hidelinks]{hyperref}
\usepackage{orcidlink}
\hypersetup{pdftitle={Where is the conformal coupling in Palatini gravity?},
pdfauthor={Alexandros Karam and Sotirios Karamitsos}}
\newcommand{\dd}{\mathrm{d}}
\newcommand{\Rc}{\mathcal R}
\newcommand{\xc}{\xi_{\mathrm c}}
\newcommand{\xp}{\xi_{\mathrm P}}
\newcommand{\xg}{\xi_g}
\newcommand{\xe}{\xi_{\mathrm{eff}}}
\newcommand{\leff}{\lambda_{\mathrm{eff}}}
\newcommand{\mpl}{M}
\newcommand{\A}{\frac{D-1}{D-2}}

\DeclareMathOperator{\sgn}{sgn}
\DeclareMathOperator{\Tr}{Tr}
\allowdisplaybreaks[1]
\begin{document}

\title{Where is the conformal coupling in Palatini gravity?}

\author{Alexandros Karam\orcidlink{0000-0002-0582-8996}}
\email{alexandros.karam@kbfi.ee}
\affiliation{National Institute of Chemical Physics and Biophysics, R\"avala 10, 10143 Tallinn, Estonia}

\author{Sotirios Karamitsos\orcidlink{0000-0002-1773-249X}}
\email{sotirios.karamitsos@ut.ee}
\affiliation{Institute of Physics, University of Tartu, W.\ Ostwaldi 1, 50411 Tartu, Estonia}

\date{\today}

\begin{abstract}
We study the conformal coupling in the torsion-free Palatini formulation with an ordinary scalar kinetic term. Tuning the curvature coupling cannot restore Weyl invariance in a Palatini theory with an ordinary scalar kinetic term because, for an independent and inert connection, the scalar-curvature term is separately Weyl-invariant, and no finite curvature coupling cancels the Weyl variation of the kinetic term. This can be demonstrated by evaluating the energy-momentum trace  on the scalar equation of motion and noting that it is explicitly independent of the Palatini coupling. Eliminating the connection, we construct a map between the curvature coupling and the effective coupling for the equivalent metric theory, and demonstrate that the scalar continues to propagate even if the reduced kinetic coefficient vanishes. We also show that it becomes a Weyl compensator with a scale-invariant potential only in the rescaled infinite-coupling limit, which in four dimensions is directly related to Palatini $\mathcal R^2$ gravity. We calculate the running of the couplings at one loop, and we find that the conformal shift in the metric-curvature coupling’s beta function generates a metric-curvature coupling on a fixed metric-affine background, whereas the independent-curvature coupling runs homogeneously. Furthermore, we follow an alternate prescription in which the connection is allowed to respond to the scalar fluctuations, which results in a different scalar Hessian and therefore different beta functions. Nonetheless, we find that both prescriptions admit the formal infrared limit $(\xi_{\mathrm{eff}},\lambda_{\mathrm{eff}})\to(1/6,0)$ through different limits of the original couplings in the case of positive perturbative quartics. Finally, we note that trajectories starting on the positive gravitational branch encounter a vanishing gravitational coefficient at a finite scale before this endpoint.

\end{abstract}

\maketitle

\section{Introduction}
\label{sec:intro}

Nonminimal scalar couplings are a prominent feature of quantum field theory in curved spacetime. Radiative corrections require scalar-curvature couplings to cancel curvature-dependent divergences. Such couplings also arise in scalar-tensor representations of $f(R)$ gravity~\cite{Sotiriou:2006hs}, where the scalar controls the effective gravitational coupling. Related auxiliary-scalar descriptions of modified Gauss--Bonnet gravity~\cite{Nojiri:2005jg} and teleparallel theories~\cite{Wright:2016ayu} involve couplings to other geometric invariants. Nonminimal couplings provide a simple framework for inflation~\cite{Bezrukov:2007ep} and, in induced-gravity models, allow for the dynamical generation of the Planck scale. For a review of Palatini gravity and its applications to inflation, see Ref.~\cite{Gialamas2023PalatiniReview}.

In the metric formulation, a scalar field is conformally coupled to gravity for a distinguished value of the curvature coupling. Assuming that the scalar potential is also scale invariant, the same value makes the improved energy-momentum tensor traceless on-shell, and the nonminimal coupling ensures that the action is invariant under Weyl transformations (i.e. local rescalings of the metric tensor). It is therefore interesting to examine how the notion of the conformal coupling changes in the Palatini formulation, where the metric and the connection are independent of one another.

The conceptual difference between the fixed value of conformal coupling in metric gravity and the general class of Weyl invariant metric-affine gravities has been noted by Bauer and Demir~\cite{Bauer:2008zj}. Without need for a distinguished value of the curvature coupling, the kinetic term in metric-affine gravity can be gauged with the aid of nonmetricity~\cite{Mikura:2020qhc,Mikura:2021ldx,deCesare:2016mml,Sauro:2022chz,Olmo:2022ops}, while the scalar-curvature term remains Weyl invariant for an arbitrary coefficient. Conformal transformations with an independent connection and the role of auxiliary scalars were also studied in~\cite{Karahan:2012yz}. When the Weyl vector is dynamical, it may acquire a mass through the Stueckelberg mechanism~\cite{Ghilencea:2020piz, Ghilencea:2020rxc,Ghilencea:2022lcl}. The geometric origin of stress-tensor improvement has been examined in~\cite{Iosifidis:2025sjx}, while frame transformations provide a complementary description of scalar--tensor dynamics~\cite{Kozak:2018vlp,Jarv:2020qqm,Postma:2014vaa,Karamitsos:2017elm}. The effects of frame transformations on strong-coupling inflationary attractors and potential reconstruction have also been studied~\cite{Karam2021PalatiniAttractors}. Related scale- and Weyl-invariant scalar--gravity models have been constructed in the Einstein--Cartan formulation, where eliminating the auxiliary torsion yields an equivalent metric theory~\cite{Karananas:2021gco}.

We study conformal coupling in the torsion-free Palatini formulation with an ordinary scalar kinetic term and no additional connection-dependent matter couplings.
One-loop corrections to nonminimally coupled quartic inflation have been studied in both the metric and Palatini formulations in effective field theory~\cite{Markkanen:2017tun}.
Eliminating the connection gives a relation between the Palatini coupling and the canonically normalized metric coupling. We examine its branches and the degeneracy of the full scalar--metric kinetic system, distinguishing the propagating dilaton at finite coupling from the compensator obtained in a rescaled infinite-coupling limit.

Going beyond the tree level, we restrict ourselves to scalar-loop calculations with the metric held fixed. In one prescription, the independent connection is also fixed; in the other, it is eliminated classically, including its response to scalar fluctuations. We derive the bulk counterterms from the scalar heat kernel on the fixed metric-affine background and compare the resulting determinant with that of the reduced action.

Models with additional ingredients, such as curvature-squared terms~\cite{Antoniadis2018PalatiniR2} or radiative symmetry breaking~\cite{Gialamas:2020snr,Gialamas2021ScaleInvariantQuadratic}, require a separate dynamical analysis. Similarly, cancellations in a particular observable, such as a leading gravitational correction to vacuum decay~\cite{Gialamas:2022gxv}, are distinct from the local symmetry considered here.

The paper is organized as follows. Section~\ref{sec:trace} compares the metric and Palatini Weyl variations and energy-momentum traces, and discusses Weyl gauging. Section~\ref{sec:reduction} eliminates the connection and examines the coupling map, scalar dynamics, and rescaled compensator limit. Section~\ref{sec:loops} derives the scalar-loop counterterms on a fixed metric-affine background. Section~\ref{sec:auxiliary} includes the auxiliary connection response and compares the two formal infrared flows. We conclude in Section~\ref{ref:summary}.

Throughout the paper, we use the mostly plus convention for the signature of the metric and impose boundary conditions that justify discarding boundary terms after integration by parts. 

\section{Weyl invariance and the energy-momentum trace}
\label{sec:trace}

We work in $D>2$ and begin with the following action, written in a unified notation: 
\begin{align}
 S_{i}=\int\dd^Dx\sqrt{-g}\left[
-\frac12\xi_{i}   \mathfrak{R}_{i}  \phi^2
 -\frac12(\partial\phi)^2
 -V(\phi)\right].
  \label{eq:actions}
\end{align}
There is no separate Einstein--Hilbert term. The trace formulas below allow an arbitrary potential; when discussing scale invariance and the induced-gravity model, we take $V(\phi)=\lambda\phi^p/4$, with $p=2D/(D-2)$. For this power-law potential and the logarithmic field redefinitions, we work locally on a patch $\phi>0$; a homogeneous extension using $|\phi|^p$ can be used on the negative-field patch. The coefficient of the gravitational term is $f=-\xi_i\phi^2$, so positive gravitational coupling requires $\xi_i<0$ in our sign convention. In Eq.~\eqref{eq:actions}, the label $i \in \{\rm m, P\}$ keeps track of whether we are working in the metric formulation or the Palatini formulation:
\begin{align}
\mathfrak{R}_{\rm m} &\equiv
R[g],
\\
\mathfrak{R}_{\rm  P } &\equiv \Rc[g,\Gamma].
\end{align}
As usual, we use $R=R[g]$ to denote the metric scalar curvature defined through the Levi-Civita connection $ \{^{\lambda}_{\mu\nu}\}_g \equiv \frac{g^{\lambda\sigma}}{2}(\partial_\mu g_{\sigma\nu} + \partial_\nu g_{\mu \sigma} -g_{\mu\nu,\sigma})$. We also use $\Rc= \Rc[g,\Gamma]$ to denote the generalized scalar curvature that depends on both the metric and the independent connection. The scalar equation of motion can also be written in a unified way:
\begin{equation}
 \Box_g\phi-\xi_{i}  \mathfrak{R}_{i} \phi-V'(\phi)=0.
 \label{eq:scalar-eom}
\end{equation}
Note that the d'Alembertian is defined with respect to the metric $g_{\mu\nu}$. The Einstein equations in the Palatini formulation are 
supplemented  with the equation $\xi_{\rm P}\nabla^{\Gamma}_{\rho} \left( \sqrt{-g} \, \phi^2 g^{\mu\nu}\right) = 0$. We use the usual definition of the energy-momentum tensor through the variational derivative $T_{\mu\nu} \equiv -2/\sqrt{-g} \ \delta S/\delta g^{\mu\nu}$.

\subsection{Metric formulation}
We first work in the metric formulation. In order to find the distinguished value of the nonminimal coupling, we first calculate the variation of the action from a finite Weyl transformation, given by
\begin{equation}
 g_{\mu\nu}\longmapsto \hat g_{\mu\nu} = e^{2\sigma(x)}g_{\mu\nu},\qquad
 \phi\longmapsto \hat \phi = e^{-\Delta\sigma(x)}\phi.
 \label{eq:weyl}
\end{equation}
Here, we have assigned a conformal weight of $\Delta = (D-2)/2$ to the field $\phi$ for consistency. We also use the formula for the transformation of the metric scalar curvature:
\begin{align}
R[e^{2\sigma}g]=e^{-2\sigma}\left[
 R[g]-2(D-1)\Box_{g}\sigma-(D-1)(D-2)(\partial\sigma)^2\right],
 \label{eq:ricci_transf_rule}
\end{align}
For homogeneous $V$ of degree $p = 2D/(D-2)$, integration by parts gives
\begin{align}
 S_{\mathrm{m}}[\hat g,\hat\phi]-S_{\mathrm{m}}[g,\phi]
 =\int\dd^Dx\sqrt{-g}\bigg[&
 \bigl(\Delta-2\xi_{\rm m} (D-1)\bigr)\phi\partial_\mu\phi\partial^\mu\sigma
 \nonumber\\[-2pt]
 &+\frac12\bigl(\xi_{\rm m} (D-1)(D-2)-\Delta^2\bigr)
 \phi^2(\partial\sigma)^2\bigg].
 \label{eq:metric-weyl}
\end{align}
Therefore, the conformal coupling in the metric formulation is 
\begin{align}
 \xc \equiv \frac{D-2}{4(D-1)},
\end{align}
as usual. 

The role of the conformal coupling is further clarified by looking at the on-shell\footnote{Here ``on shell'' means imposing only the scalar equation of motion. If the metric is also varied, and no other sectors are present, its equation is $T_{\mu\nu}=0$.} trace of the energy-momentum tensor. The off-shell energy-momentum tensor is
\begin{equation}
 T^{\mathrm{m}}_{\mu\nu}
 =\partial_\mu\phi\partial_\nu\phi-\frac12g_{\mu\nu}(\partial\phi)^2
 -g_{\mu\nu} V
 +\xi_{\rm m}\left[G_{\mu\nu}\phi^2+
 (g_{\mu\nu}\Box_g-\nabla_\mu\nabla_\nu)\phi^2\right].
 \label{eq:Tmetric}
\end{equation}
The derivative term supplies the familiar improvement. Using
Eq.~\eqref{eq:scalar-eom}, its trace takes the compact form
\begin{equation}
 T^{\mathrm{m}}\big|_{\text{on-shell}}
 =(D-1)(\xi_{\rm m}-\xc)\Box_g\phi^2+\Delta\phi V'-DV.
 \label{eq:metric-trace}
\end{equation}
For a scale-invariant potential, such as $V=\lambda\phi^p/4$, the trace vanishes on every scalar solution when the coupling takes on its conformal value $\xi_{\rm m}=\xc$. For other values of the coupling, the trace can vanish for particular field configurations, but it is identically zero on all scalar solutions only in the case of conformal coupling.

\subsection{Palatini formulation}

We now turn to the Palatini formulation to examine the variation of the action. We take the independent connection $\Gamma^\rho{}_{\mu\nu}$ to be inert. The generalized curvature then transforms homogeneously, $\Rc\mapsto e^{-2\sigma}\Rc$. Thus, $\sqrt{-g}\,\Rc\phi^2$ is Weyl invariant separately, at every coupling. The entire variation comes from the ordinary kinetic term:
\begin{equation}
 S_{\mathrm P}[\hat g,\hat\phi,\Gamma]-S_{\mathrm P}[g,\phi,\Gamma]
 =\int\dd^Dx\sqrt{-g}\left[
 \Delta\phi\partial_\mu\phi\partial^\mu\sigma
 -\frac12\Delta^2\phi^2(\partial\sigma)^2\right].
 \label{eq:palatini-weyl}
\end{equation}
For $D>2$, no finite choice of $\xp$ cancels the variation. 

We arrive at a similar conclusion when looking at the trace. The Palatini energy-momentum tensor is instead
\begin{equation}
 T^{\mathrm P}_{\mu\nu}
 =\partial_\mu\phi\partial_\nu\phi-\frac12g_{\mu\nu}(\partial\phi)^2
 -g_{\mu\nu}V
 +\xp\phi^2\left[R_{(\mu\nu)}(\Gamma)-\frac12g_{\mu\nu}\Rc\right],
 \label{eq:Tpalatini}
\end{equation}
where the symmetrization in $R_{(\mu\nu)}$ includes a factor of $1/2$.
Here, there is no derivative improvement term. On-shell, we obtain
\begin{equation}
 T^{\mathrm{P}}\big|_{\text{on-shell}}
 =-\frac{D-2}{4}\Box_g\phi^2+\Delta\phi V'-DV
 =-\frac{D-2}{4}\Box_g\phi^2,
 \label{eq:palatini-trace}
\end{equation}
for homogeneous $V$. This expression has no explicit dependence on $\xp$ and has the same form as the metric trace at $\xi_{\rm m}=0$ in Eq.~\eqref{eq:metric-trace}. The scalar solutions themselves still depend on $\xp$. 
Note that in flat space with $\Gamma^\rho{}_{\mu\nu}=0$, the two scalar actions coincide. The virial current in Eq.~\eqref{eq:palatini-trace} is itself a derivative, and the flat-space stress tensor can be improved in the usual way (adding a divergenceless term to the stress tensor so that its on-shell trace vanishes)~\cite{Callan:1970ze, Iorio:1996ad}. In general, metric-affine Ward identities can involve hypermomentum~\cite{Iosifidis:2025sjx}, but they depend on the specifics of the matter couplings and the transformation rule for the curvature. With an inert independent connection, there is no connection-variation term in the Weyl variation of Eq.~\eqref{eq:actions}, and so an obstacle arises in this particular curved-space Weyl completion.

\subsection{Weyl gauging the scalar kinetic term}
\label{sec:gauging}

Modifying the kinetic term of the action is a way to achieve Weyl invariance, but this does not distinguish a particular value of the coupling as the conformal coupling. Defining nonmetricity as
\begin{equation}
 Q_{\lambda\mu\nu}=-\nabla^\Gamma_\lambda g_{\mu\nu},\qquad
 Q_\lambda=g^{\mu\nu}Q_{\lambda\mu\nu},
 \label{eq:Q}
\end{equation}
it transforms as $Q_\mu\mapsto Q_\mu-2D\partial_\mu\sigma$ if the connection is kept inert. This allows us to define the Weyl-covariant derivative~\cite{Mikura:2020qhc,Sauro:2022chz}:
\begin{equation}
 \mathscr D_\mu\phi=(\partial_\mu+\Delta W_\mu)\phi
 =\left(\partial_\mu-\frac{D-2}{4D}Q_\mu\right)\phi,
 \label{eq:weyl-derivative}
\end{equation}
where $ W_\mu \equiv -\frac{Q_\mu}{2D}$. This definition agrees with~\cite{Mikura:2020qhc} for $D=4$.

The Weyl-covariant derivative transforms as  $\mathscr D_\mu\phi\mapsto e^{-\Delta\sigma}
\mathscr D_\mu\phi$. Replacing $\partial_\mu$ by $\mathscr D_\mu$ makes the kinetic term Weyl invariant in isolation for any value of the coupling $\xp$. The resulting action, which is manifestly Weyl invariant, now leads to a different connection equation, since it appears implicitly in the kinetic term through $W_\mu$. Since $W_\mu=-Q_\mu/(2D)$, its variation must be included when varying the connection. Treating $W_\mu$ as an independent variable defines a different variational problem. If suitable kinetic terms for $W_\mu$ are introduced, the scalar can instead supply its longitudinal mode via a Stueckelberg mechanism~\cite{Ghilencea:2020piz,Ghilencea:2020rxc,Ghilencea:2022lcl}.

\section{Classical reduction and scalar dynamics}
\label{sec:reduction}

So far, we have treated the connection and metric as a background. In this section, we return to an ordinary kinetic term and, working in the Palatini formulation, we treat both the metric and the connection as dynamical. After eliminating the independent connection, we will be able to compare the couplings of a Palatini theory with those of a metric theory. 

\subsection{Exact reduction by eliminating the independent connection }

Our first goal is to eliminate the independent connection and express the theory entirely in terms of the metric and scalar. We first write the gravitational sector for a general coefficient $f(\phi)$:
\begin{align}
S_{\rm grav}=\frac12\int\dd^Dx\sqrt{-g}\,f(\phi)\Rc.
\end{align}
The scalar kinetic term and potential are independent of the connection, so its equation of motion is 
\begin{equation}
 \nabla^\Gamma_\lambda\left(\sqrt{-g}\,f g^{\mu\nu}\right)=0.
 \label{eq:compat}
\end{equation}
For a symmetric connection in $D>2$, the unique solution on a fixed-sign patch with $f\ne0$ is the Levi-Civita connection of the conformally related metric
\begin{equation}
 h_{\mu\nu}=\left(\frac{|f|}{\mpl^{D-2}}\right)^{2/(D-2)}g_{\mu\nu},
 \label{eq:hmetric}
\end{equation}
where the constant $\mpl$ fixes units. For the $f(\phi)\Rc$ action with connection-independent matter, an unrestricted connection has an arbitrary projective vector in its solution~\cite{Olmo:2011uz}. Imposing zero torsion selects the torsion-free representative and removes that freedom.

In general, an ordinary projective shift of a general connection does not preserve the symmetry of its lower indices. Instead, its symmetrized version does preserve the absence of torsion, but also changes the curvature scalar (where $V_\mu$ is an arbitrary smooth one-form):
\begin{align}
 \Gamma^\lambda{}_{\mu\nu}&\longmapsto
 \Gamma^\lambda{}_{\mu\nu}+\delta^\lambda{}_{(\mu}V_{\nu)},\nonumber
 \\
 \Rc&\longmapsto\Rc-\frac{D-1}{2}g^{\mu\nu}
 \nabla^\Gamma_\mu V_\nu+\frac{D-1}{4}V_\mu V^\mu.
 \label{eq:symmetric-shift}
\end{align}
Since the connection is not metric compatible, the contraction $g^{\mu\nu}\nabla^\Gamma_\mu V_\nu$ cannot in general be replaced by $\nabla^\Gamma_\mu V^\mu$. Since this shift contains contributions beyond a boundary term, it is not a gauge freedom of the action. 

Relating the generalized curvature scalar with the one derived from the metric yields
\begin{align}
 \Rc&=R[g]-\frac{2(D-1)}{D-2}\Box_g\ln|f|
 -\A(\partial\ln|f|)^2,\label{eq:R-reduced}
\end{align}
which allows us to write the reduced action as
\begin{align}
 S_{\mathrm{red}}&=\int\dd^Dx\sqrt{-g}\left[
 \frac12 fR[g]-\frac12 k_{\mathrm{red}}(\partial\phi)^2-V\right],
 \qquad
 k_{\mathrm{red}}=k-\A\frac{(f')^2}{f}.
 \label{eq:general-reduction}
\end{align}
We have introduced a more general notation for the original kinetic term, which is $k=1$ for Eq.~\eqref{eq:actions}. For $f=-\xp\phi^2$, we find that the reduced action becomes
\begin{equation}
 S_{\mathrm{red}}=\int\dd^Dx\sqrt{-g}\left[
 -\frac12  Z  (\partial\phi)^2-\frac12\xp R[g]\phi^2-V\right],
\end{equation}
where we define the kinetic coefficient
\begin{align}
Z \equiv  1+\frac{\xp}{\xc}.
 \label{eq:Z}
\end{align}
Thus, eliminating the connection has generated a modification to the kinetic term. These classical actions are equal up to a boundary term. In more general models, the reduction is still possible but the kinetic term is field-dependent~\cite{Bauer:2008zj,Kozak:2018vlp}. The variation of the connection is written out in full in Appendix~\ref{app:connection}.

\subsection{The coupling map and its branches}

For a positive $Z$, rescaling the field as  $\varphi=\sqrt Z\,\phi$ returns a metric theory with a canonical kinetic term and the following effective nonminimal coupling and potential coupling:  
\begin{equation}
 \xe=\frac{\xp}{1+\xp/\xc},\qquad
 \leff=\frac{\lambda}{Z^{D/(D-2)}}.
 \label{eq:harmonic}
\end{equation}
Assuming $\xp\ne0$, we see that the effective coupling is related to the original Palatini coupling and the conformal coupling by the inverse-coupling relation, $ \frac1{\xe}=\frac1{\xp}+\frac1{\xc}$. The rescaling is invertible on its regular branch and does not fix $\xp$. Indeed, in these variables, the kinetic contribution from the connection is absorbed into the curvature and potential couplings, in agreement with
the frame description of Refs.~\cite{Kozak:2018vlp,Jarv:2020qqm}. Hence, tuning $\xp$ cannot turn the scalar into a Weyl compensator for dynamical gravity.

\begin{figure}[tb]
\centering
\begin{minipage}[t]{0.48\textwidth}
    \centering
    {\small \hspace{2.5em} (a) Coupling map }
    \smallskip
    \includegraphics[width=\linewidth]{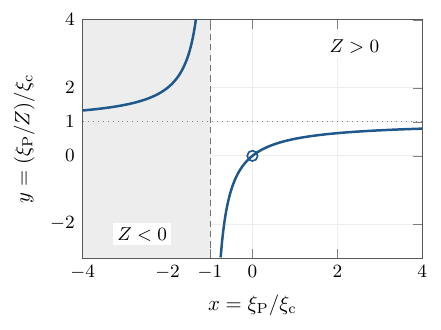}
\end{minipage}
\hfill
\begin{minipage}[t]{0.48\textwidth}
    \centering
    {\small \hspace{2.5em} (b) Kinetic degeneracy }
    \smallskip
    \includegraphics[width=\linewidth]{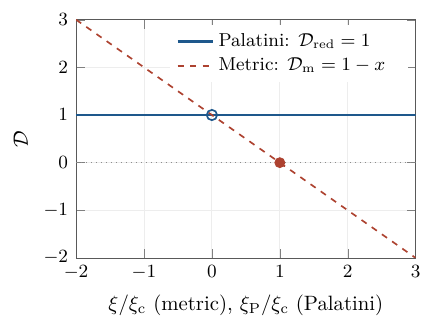}
\end{minipage}
\caption{The coupling map and kinetic degeneracy in  conformally normalized  and dimensionless variables, valid for any $D>2$. (a) The ratio $y=x/(1+x)$ has a pole at $x=-1$ and approaches $y=1$ as $|x|\to\infty$. In the shaded region $Z<0$, $y$ is a ratio of coefficients, not a canonically normalized positive-kinetic coupling. (b) For $k=1$, the metric kinetic combination vanishes at $\xc$, while the reduced Palatini combination remains unity. The open circles mark $f=0$, which is excluded from the reduction process. The sign of $Z$ alone does not determine the physical scalar kinetic sign: every finite $\xp<0$ has $\mathcal K_{\mathrm P}>0$.}
\label{fig:branches}
\end{figure}

We note that Eq.~\eqref{eq:harmonic} is not a global map between metric theories, since at $\xp=-\xc$,  the rescaling is singular. For $Z<0$, a normalization $\varphi=\sqrt{|Z|}\phi$ instead gives
\begin{equation}
 k_\varphi=\sgn Z,\qquad
 \xi_\varphi=\frac{\xp}{|Z|},\qquad
 \lambda_\varphi=\frac{\lambda}{|Z|^{D/(D-2)}}.
 \label{eq:real-normalization}
\end{equation}
One can still define the algebraic ratio $\xp/Z$, but it is then $\xi_\varphi/k_\varphi$, not the coupling of a scalar with $k_\varphi=1$. 
The different branches are shown in Fig.~\ref{fig:branches}. For $\xp>0$, one has $f<0$ and $0<\xe<\xc$. For $-\xc<\xp<0$, both $f$ and $Z$ are positive and $\xe<0$. For $\xp<-\xc$, $f>0$ but $Z<0$, and the algebraic ratio $\xp/Z$ exceeds $\xc$. Therefore, the bound $\xe<\xc$ applies to the canonical branch $Z>0$ and cannot be extended to all real Palatini couplings.

\subsection{Scalar--metric kinetic degeneracy}

Consider a metric scalar-tensor action of the form given in Eq.~\eqref{eq:general-reduction}, with nonminimal coefficient $f(\phi)$, metric kinetic coefficient $k_{\rm m}(\phi)$, and potential $V(\phi)$. On a regular fixed-sign patch, a Weyl transformation to the Einstein frame gives the scalar kinetic coefficient
\begin{align}
 \mathcal K_{\mathrm{m}}
 &=\frac{\mpl^{D-2}}{|f|}\mathcal D_{\mathrm{m}},  \nonumber
 \\
\mathcal D_{\mathrm{m}} &\equiv  k_{\rm m} +\A \frac{f'^2}{f}  ,
 \label{eq:Kmetric}
\end{align}
where we have introduced the dimensionless quantity $\mathcal D_{\mathrm m}$. On the positive gravitational branch, the field-space coefficient of the invariant scalar--tensor formulation is $\mathcal F=\mathcal D_{\mathrm m}/(2f)$~\cite{Jarv2015ScalarTensorInvariants,Karam2018FrameInvariants}. The condition $\mathcal D_{\mathrm m}=0$ is preserved by regular frame transformations and invertible field reparametrizations. The kinetic degeneracy can also occur on the $f<0$ branch, where the Einstein--Hilbert coefficient has the opposite sign.

For the standard metric model with a canonical kinetic term $k_{\rm m}=1$ and a nonminimal coupling $f=-\xi_{\rm m}\phi^2$, we find $\mathcal D_{\mathrm m}=1-\xi_{\rm m}/\xc$, so the degeneracy occurs at $\xi_{\rm m}=\xc$. With a homogeneous potential of degree $2D/(D-2)$, this degeneracy precisely corresponds to a local Weyl symmetry: on a nonzero-field patch, the scalar acts as a compensator (and can be gauge fixed). In this sign convention, this particular model has $f<0$. A compensator action with positive gravitational coupling instead has a negative Jordan-frame scalar kinetic coefficient. The scalar is a gauge variable, so its negative Jordan-frame kinetic coefficient does not imply an additional physical ghost.

We now turn to the reduced Palatini model. The connection-induced kinetic contribution cancels the contribution from the transformation to Einstein-frame variables. Substituting $k_{\rm m}=k_{\rm red}$ from Eq.~\eqref{eq:general-reduction}, we obtain
\begin{align}
 \mathcal D_{\mathrm{red}}
 &=k_{\rm red}+\A\frac{(f')^2}{f}\nonumber\\
 &=\left(k-\A\frac{(f')^2}{f}\right)
   +\A\frac{(f')^2}{f}=k.
 \label{eq:Dpalatini}
\end{align}
For the original Palatini model, $k=1$, so $\mathcal D_{\mathrm{red}}=1$ and the scalar--metric kinetic system remains nondegenerate at every finite nonzero $\xp$. In particular, $Z=0$ at $\xp=-\xc$ is not a conformal point: the scalar dynamics comes entirely from scalar--metric mixing. On the positive gravitational branch, $f>0$ and hence $|f|=f$, so the Einstein-frame coefficient is
\begin{equation}
 \mathcal K_{\mathrm P}(\phi)=\frac{\mpl^{D-2}}{f}
 =-\frac{\mpl^{D-2}}{\xp\phi^2}>0,\qquad \xp<0.
 \label{eq:Kpalatini}
\end{equation}

In four dimensions, the point $Z=0$ gives
\[
 S_{\mathrm{red}}=\int\dd^4x\sqrt{-g}
 \left[\frac{\phi^2R[g]}{12}-\frac\lambda4\phi^4\right].
\]
For $\lambda\ne0$ and on a nonzero real branch, the scalar equation gives $\phi^2=R[g]/(6\lambda)>0$. Eliminating $\phi$ then yields
\[
 S_{\mathrm{red}}=\frac1{144\lambda}\int\dd^4x\sqrt{-g}\,R[g]^2.
\]
This is metric $R^2$ gravity, which retains a scalar degree of freedom through scalar--metric mixing, in contrast to the Palatini $\Rc^2$ limit discussed below~\cite{Edery:2019txq}.\footnote{On $f>0$, setting $\Phi=f=-\xp\phi^2$ gives a Brans--Dicke action with $\omega_{\rm BD}=-(D-1)/(D-2)-1/(4\xp)$. The point $Z=0$ has $\omega_{\rm BD}=0$, whereas the critical value $-(D-1)/(D-2)$ is reached only as $|\xp|\to\infty$; see Ref.~\cite{Olmo:2011uz} for the Palatini $f(R)$ correspondence.}

The different behavior of the two kinetic combinations is illustrated in Fig.~\ref{fig:branches}(b).

\subsection{Finite-coupling dilaton and the Weyl-invariant limit}

For finite $\xp<0$ and $\phi>0$, the action can be cast in the canonical Einstein frame by reparametrizing the field as follows:
\begin{equation}
 \chi=\frac{\mpl^{(D-2)/2}}{\sqrt{-\xp}}\ln\frac{\phi}{v},\qquad
 U(\phi)=\left(\frac{\mpl^{D-2}}{f}\right)^{D/(D-2)}V(\phi)
 =\frac{\lambda\mpl^D}{4(-\xp)^{D/(D-2)}}\equiv U_0.
 \label{eq:dilaton}
\end{equation}
Here $v>0$ is an arbitrary reference value. Consequently, the action becomes
\begin{equation}
 S_E=\int\dd^Dx\sqrt{-g_E}\left[
 \frac{\mpl^{D-2}}2R[g_E]-\frac12(\partial\chi)^2-U_0\right].
 \label{eq:einstein-dilaton}
\end{equation}
This action describes a model that contains Einstein gravity, a minimally coupled massless scalar, and a cosmological constant $\Lambda=U_0/\mpl^{D-2}$. The Einstein-frame scalar has a constant potential.  

The ratio $\xp/Z$ tends to $\xc$ as $|\xp|\to\infty$, but this does not necessarily represent a loss of a degree of freedom.  We write $\xp=s\rho$, where $s=\pm1$ and $\rho>0$, and introduce the rescaled quantities
\begin{equation}
 \psi=\sqrt\rho\,\phi,\qquad
 \bar\lambda=\lambda\rho^{-D/(D-2)}.
 \label{eq:limit-rescaling}
\end{equation}
With these definitions, the action becomes
\begin{equation}
 S_{\mathrm P}=\int\dd^Dx\sqrt{-g}\left[
 -\frac{s}{2}\psi^2\Rc-\frac{1}{2\rho}(\partial\psi)^2
 -\frac{\bar\lambda}{4}\psi^p\right].
 \label{eq:limit-action}
\end{equation}
At fixed $\psi$ and $\bar\lambda$, the $\rho\to\infty$ limit has no ordinary scalar kinetic term and is indeed locally Weyl invariant with an inert connection. Eliminating the connection gives $f_\psi=-s\psi^2$ and $k_\psi=s/\xc$, so $\mathcal D_{\mathrm{red}}=0$ as expected. In this limiting theory, $\psi$ acts as a Weyl compensator: on a nonzero-field patch, the local Weyl symmetry allows it to be fixed to a constant, yielding Einstein gravity with a constant potential. Its gravitational coupling is positive for $s=-1$.
More explicitly, the bare kinetic coefficient in these variables is $k=1/\rho$, which leads to $\mathcal D_{\mathrm{red}}=1/\rho$ before taking the limit. The degeneracy therefore arises only at the endpoint. If the original $\lambda$ is kept fixed, the limiting potential vanishes. The limiting theory with $1/\rho = 0$ is therefore well-defined after a rescaling, but it is a singular limit of the finite-coupling theories: the change to a canonically normalized propagating scalar becomes singular there. No scalar is removed at any finite $\rho$, and the  proximity of a coupling ratio to $\xc$ alone is not a criterion for treating deviations from a Weyl-invariant theory as small.

In four dimensions, the limiting theory also has a direct relation to Palatini $\Rc^2$ gravity~\cite{Edery:2019txq}. For $\bar\lambda\ne0$ and on a nonzero real branch, the auxiliary scalar equation gives $\psi^2=-s\Rc/\bar\lambda$. Substituting into the limiting action yields
\begin{equation}
 S_\infty=\frac{1}{4\bar\lambda}\int\dd^4x\sqrt{-g}\,\Rc^2.
 \label{eq:R2-endpoint}
\end{equation}
This is the homogeneous curvature action selected by Weyl invariance within the minimal four-dimensional Palatini $f(\Rc)$ sector, as also discussed in Ref.~\cite{Coumbe:2026zdm}. The elimination requires $-s\Rc/\bar\lambda>0$ on the chosen patch. It does not apply when $\bar\lambda=0$, where the nonzero-scalar equation instead imposes $\Rc=0$.

\section{Scalar loops on a fixed metric-affine background}
\label{sec:loops}

In this section, we demonstrate the effects of the conformal coupling (or lack thereof) on loop calculations. We assume $D=4$ and we work with dimensional regularization  $d = 4-\epsilon$ in the minimal subtraction scheme. We treat gravity (both the metric and the connection) as a background, but without assuming that the background connection satisfies its classical equation of motion.

\subsection{Scalar fluctuation operator and beta functions}

To capture the effects of renormalization, we must include both curvature couplings \cite{Brown:1980qq}. After a Wick rotation, the scalar action becomes
\begin{equation}
 S_E^{\mathrm{bg}}=\int\dd^4x\sqrt g\left[
 \frac12(\partial\phi)^2+\frac12(\xp\Rc+\xg R)\phi^2
 +\frac\lambda4\phi^4\right],\qquad R=R[g].
 \label{eq:background-action}
\end{equation}
Separating the fluctuation from the scalar through $\phi=\bar\phi+\eta$, we find the scalar operator to be 
\begin{equation}
 \mathscr P=-\Box_g+E,\qquad
 E=3\lambda\bar\phi^2+\xp\Rc+\xg R.
 \label{eq:operator}
\end{equation}
The independent connection appears only in $E$ through $\Rc$.

We now expand the heat-kernel trace in order to define the local coefficients as
\begin{align}
\Tr e^{-s\mathscr P}\sim(4\pi s)^{-2}
\int \dd^4 x \, \sqrt g\,[a_0 \, + \,s a_2+s^2a_4 \, + \, \cdots].
\end{align}
The standard heat kernel approach~\cite{Vassilevich:2003xt} gives
\begin{equation}
 a_4=\frac12\left(E-\frac R6\right)^2
 +\frac1{180}\left(R_{\mu\nu\rho\sigma}R^{\mu\nu\rho\sigma}
 -R_{\mu\nu}R^{\mu\nu}\right)
 +\frac1{30}\Box_g R-\frac16\Box_g E.
 \label{eq:a4}
\end{equation}
Again, save for the terms in $\Rc$, all terms are standard Levi-Civita quantities. With dimensional regularization, the Euclidean counterterm that cancels the pole is
\begin{equation}
 S_{\mathrm{ct}}=\frac{1}{16\pi^2\epsilon}
 \int\dd^4x\sqrt g\,a_4.
 \label{eq:counterterm}
\end{equation}
The normalization of the pole is fixed by $d= 4-\epsilon$: using $d=4-2\epsilon$ instead
would change the denominator of the coefficient to $32\pi^2\epsilon$.

The scalar-dependent bulk terms in the $a_4$ coefficient are given by
\begin{equation}
 a_4\supset \frac92\lambda^2\bar\phi^4
 +3\lambda\bar\phi^2
 \left[\xp\Rc+\left(\xg-\frac16\right)R\right].
 \label{eq:matter-counterterms}
\end{equation}
There is no one-loop bulk wave-function counterterm in this model. The beta functions, with $t=\ln(\mu/\mu_0)$, are therefore
\begin{equation}
 \beta_\lambda=\frac{9\lambda^2}{8\pi^2},\qquad
 \beta_{\xp}=\frac{3\lambda}{8\pi^2}\xp,\qquad
 \beta_{\xg}=\frac{3\lambda}{8\pi^2}
 \left(\xg-\frac16\right).
 \label{eq:betas}
\end{equation}
For $\xp=0$, these beta functions correspond to the standard results in the normalization $V=\lambda\phi^4/4$~\cite{Markkanen:2013nwa,Merzlikin:2017zan}. 

Starting with a pure Palatini coupling,  $\xg=0$, a metric coupling is still generated because $\beta_{\xg}=-\lambda/(16\pi^2)$ at $\xg=0$. The homogeneous running of $\xp$ follows from the role of $\Rc$ as a potential in the Laplace-type operator. 

\subsection{Curvature counterterms and nonmetricity}

Eqs.~\eqref{eq:a4} and \eqref{eq:matter-counterterms} also determine the complete set of required one-loop counterterms for the operator in Eq.~\eqref{eq:operator}.  The scalar loop generates divergences proportional to $\phi^2\Rc$, $\phi^2R$, and $\phi^4$, as well as terms involving only the background geometry. Up to total derivatives, these purely geometric divergences are cancelled by counterterms proportional to
\begin{equation}
 \Rc^2,\quad \Rc R,\quad R^2,\quad C_{\mu\nu\rho\sigma}C^{\mu\nu\rho\sigma},
 \quad E_4,
 \label{eq:vacuum-basis}
\end{equation}
where $E_4=R_{\mu\nu\rho\sigma}^2-4R_{\mu\nu}^2+R^2$. The non-metricity is found in $\Rc$ and its square or product with $R$, together with the total derivative $\Box_g\Rc$.

We can explicitly see that the non-metricity dependence is contained in $\Rc$: for $N^\lambda{}_{\mu\nu}=\Gamma^\lambda{}_{\mu\nu}
-\{^{\lambda}_{\mu\nu}\}_g$, one has
\begin{equation}
 \Rc-R=g^{\mu\nu}\left[
 \nabla^{(g)}_\lambda N^\lambda{}_{\mu\nu}
 -\nabla^{(g)}_\nu N^\lambda{}_{\mu\lambda}
 +N^\lambda{}_{\lambda\rho}N^\rho{}_{\mu\nu}
 -N^\lambda{}_{\nu\rho}N^\rho{}_{\mu\lambda}\right].
 \label{eq:distortion}
\end{equation}
For a torsion-free connection, $N$ is determined algebraically by non-metricity. By expanding Eq.~\eqref{eq:distortion} in terms of non-metricity invariants, we can see that the $\phi^2Q^2$ and $\phi^2\nabla Q$ terms appear with the relative coefficients fixed by the decomposition of $\phi^2\Rc$. Instead, a larger basis including every allowed $\phi^2Q^2$ or $\phi^2\nabla Q$ operator becomes relevant for more general scalar operators, additional matter couplings, or quantum connection fluctuations. 

Equivalently, we may change basis to separate the contribution of the independent connection. If $\xi_R=\xp+\xg$, then we rewrite $ \xp\Rc+\xg R=\xi_RR+\xp(\Rc-R)$, and the beta function is
\begin{equation}
 \beta_{\xi_R}=\frac{3\lambda}{8\pi^2}
 \left(\xi_R-\frac16\right).
 \label{eq:distortion-running}
\end{equation}
The result demonstrates the familiar metric conformal shift and homogeneous running of the distortion coupling. Similar constructions which separate metric-curvature and non-Riemannian background couplings can be performed in matter theories with torsion~\cite{Shapiro:2001rz} and external two-forms~\cite{Buchbinder:2025drp}, while general metric-affine heat-kernel and anomaly calculations provide a wider setting~\cite{Bahamonde:2024kyi}. Nonetheless, promoting the backgrounds to quantum gravitational fields is a considerably more complicated problem that requires the construction of a gravitational effective action \cite{Pradisi:2022nmh,Baldazzi:2021kaf}.

\subsection{Infrared flow towards the metric conformal coupling}

We can now examine whether quantum running drives the curvature couplings towards particular values as the energy scale decreases. For $\lambda_0>0$, the perturbative solutions for the running couplings are
\begin{align}
 \lambda(t)&=\frac{\lambda_0}{1-9\lambda_0t/(8\pi^2)},\qquad
 q(t)=\left[\frac{\lambda(t)}{\lambda_0}\right]^{1/3},\nonumber\\
 \xp(t)&=\xp(0)q(t),\qquad
 \xg(t)=\frac16+\left(\xg(0)-\frac16\right)q(t).
 \label{eq:RGsolution}
\end{align}
Along these RG flow trajectories and under this massless one-loop approximation, $t\to-\infty$ gives $(\lambda,\xp,\xg)\to(0,0,1/6)$. This formal minimal-subtraction extrapolation does not include decoupling at curvature- or field-dependent mass thresholds. At $\lambda=0$, all values of $(\xp,\xg)$ have vanishing beta functions in Eq.~\eqref{eq:betas}. At nonzero $\lambda$, however, $(\xp,\xg)=(0,1/6)$ is the only simultaneous zero of the two curvature beta functions, and it is preserved while $\lambda$ runs. It is not an interacting fixed point of the complete matter-coupling flow, since $\beta_\lambda\ne0$. 

The point $(0,1/6)$ describes a classically conformally coupled metric scalar
on the prescribed background, but the running towards it does not mean that an interacting
theory dynamically restores local Weyl gauge symmetry. Indeed, even a free conformally coupled real scalar features a gravitational trace anomaly:
\begin{equation}
 \langle T^\mu{}_{\mu}\rangle
 =\frac1{(4\pi)^2}\left[
 \frac1{120}C_{\mu\nu\rho\sigma}^2-\frac1{360}E_4
 +b\Box_gR\right],
 \label{eq:anomaly}
\end{equation}
where $b$ depends on the finite $R^2$ counterterm~\cite{Vassilevich:2003xt}.
Interactions add beta-function terms, and quantum running selects the metric conformal coupling, even when the starting action includes a Palatini curvature coupling.

\section{Scalar loops with an auxiliary connection}
\label{sec:auxiliary}

In the previous section, we treated the connection as a fixed background. We will now adopt an alternative prescription in order to examine how the scalar-loop calculation changes when the connection responds to scalar fluctuations. To do so, we still hold the metric fixed, but eliminate the connection using its classical equation of motion, $\Gamma=\Gamma_\star[g,\phi]$, and express fluctuations of the connection in terms of the scalar fluctuations. 

\subsection{The reduced fluctuation operator and beta functions}

We follow an alternative prescription in which we eliminate the connection and then quantize the reduced action at fixed $g$. In four dimensions, with $\xg=0$ and on a regular patch $\xp\ne0$, $Z>0$, the Euclidean scalar Hessian is
\begin{equation}
 \mathscr H_{\mathrm{red}}=-Z\Box_g+\xp R+3\lambda\bar\phi^2.
 \label{eq:reduced-Hessian}
\end{equation}
After writing $\varphi=\sqrt Z\,\phi$, the canonical operator is
\begin{equation}
 \mathscr P_{\mathrm{red}}=-\Box_g+\xe R+3\leff\bar\varphi^2,
 \qquad \xe=\frac{\xp}{Z},\quad \leff=\frac{\lambda}{Z^2}.
 \label{eq:reduced-operator}
\end{equation}
Therefore, the operator has the standard metric scalar beta functions in
$(\xe,\leff)$. This conclusion follows directly from the reduced action;
one must vary the solution $\Gamma_\star[g,\phi]$ when taking its second
scalar variation.

For clarity, the corresponding pullback to the local coordinates
$(\xp,\lambda)$ is
\begin{equation}
 \beta^{\mathrm{red}}_{\xp}
 =-\frac{3\leff}{8\pi^2}\left(\xp+\frac16\right),\qquad
 \beta^{\mathrm{red}}_\lambda
 =\frac{3\lambda^2}{8\pi^2 Z^2}.
 \label{eq:reduced-betas}
\end{equation}
The second equation includes the running of $Z$ and is essential to the change of coordinates. The pullback also fixes the running field normalization through $\varphi=\sqrt Z\,\phi$. Although the canonical scalar has no one-loop wavefunction divergence, the original field coordinate inherits
\[
 \gamma_\phi=\frac12\frac{\dd\ln Z}{\dd t}
 =-\frac{3\leff}{16\pi^2},\qquad
 \mu\frac{\dd\phi}{\dd\mu}=-\gamma_\phi\phi.
\]
This induced running explains the quartic coefficient in Eq.~\eqref{eq:reduced-betas}; it does not represent an additional scalar self-energy divergence. These formulas are valid only where the coordinate map and scalar determinant are regular. In particular, $\xp=-1/6$ lies outside this domain, at which point we have $Z=0$, so the canonical
normalization breaks down and $\leff=\lambda/Z^2$ is undefined. The apparent factor $\xp+1/6$ in the first beta function therefore  does not establish a fixed point: the metric conformal value $\xe=1/6$ has no finite preimage under the coupling map.

\subsection{Scalar--connection mixing and the one-loop determinant}

We can now see why the two prescriptions produce different beta functions. The difference between Eqs.~\eqref{eq:betas} and \eqref{eq:reduced-betas} follows from their fluctuation operators. Quantizing a scalar at fixed $\Gamma$ requires us to use $S_{,\phi\phi}$, whereas to eliminate an auxiliary connection we must use the reduced Hessian given by
\begin{equation}
 S_{\mathrm{red},\phi\phi}
 =S_{,\phi\phi}
 -S_{,\phi\Gamma}(S_{,\Gamma\Gamma})^{-1}S_{,\Gamma\phi},
 \label{eq:schur}
\end{equation}
evaluated at $\Gamma_\star$. This expression includes the mixed fluctuation term $S_{,\Gamma \phi}$ which is absent from the fixed-connection determinant. Crucially, this term cannot be supplied by setting the background connection on shell after computing the fixed-connection determinant: it must be taken into account from the beginning. In Appendix~
\ref{app:hessian}, we derive Eq.~\eqref{eq:schur} and verify Eq.~\eqref{eq:reduced-operator}.

We can also illustrate the difference between the two prescriptions by examining the path integral. Integrating over the auxiliary fluctuations returns the block-determinant identity
\begin{equation}
 \det S^{(2)}
 =\det S_{,\Gamma\Gamma}\;\det S_{\mathrm{red},\phi\phi},
 \label{eq:blockdet}
\end{equation}
on a regular patch with any gauge zero modes handled appropriately. The auxiliary determinant and measure must be treated consistently and cannot simply be dropped in an arbitrary regulator. The comparison here concerns two choices of fluctuating fields: a scalar at fixed $g$ and $\Gamma$, yielding Eq.~\eqref{eq:betas}, and scalar quantization after classically eliminating $\Gamma$ at fixed $g$, yielding Eq.~\eqref{eq:reduced-betas}. In the latter prescription, the response of the connection is incorporated into the scalar Hessian. Changing the order of integration in a single consistently defined path integral does not, by itself, define a different quantum theory. Frame-covariant treatments of quantum scalar--tensor theories also emphasize the role of the functional measure~\cite{Finn2020QuantumFrameCovariance}; their inclusion of metric fluctuations goes beyond the scalar fluctuation problems considered here.

\subsection{Infrared limits of the effective couplings}

The two prescriptions lead to different beta functions, but admit a common formal limit for the canonically normalized curvature coupling. We choose the same initial couplings $\xp(0)=P_0\ne0$, $\xg(0)=0$, and $\lambda(0)=\lambda_0>0$, with $Z_0=1+P_0/c>0$. We also assume that both $\lambda_0$ and $\lambda_0/Z_0^2$ are perturbative, and define
\begin{equation}
 q_A=(1-a\lambda_0t)^{-1/3},\qquad
 q_B=(1-a\lambda_0t/Z_0^2)^{-1/3},
 \label{eq:IR-q}
\end{equation}
where we have set the four-dimensional conformal coupling $c\equiv 1/6$ and the coefficient of the quartic beta function as $a\equiv9/(8 \pi^2)$ for brevity. 
Here we distinguish the prescriptions: $A$ denotes the independent-background calculation and $B$ the calculation in the reduced theory. 

In prescription $A$, we formally reduce the two-derivative scalar-curvature sector at each scale, leaving aside the generated background curvature-squared operators. This gives $Z_A=1+\xp^A/c$ and the total metric curvature coefficient
$\xp^A+\xg^A$. Hence
\begin{align}
 \xp^A&=P_0q_A,\qquad \xg^A=c(1-q_A),\nonumber\\
 \xe^A-c&=\frac{\xp^A+\xg^A}{1+\xp^A/c}-c
 =-\frac{c q_A}{1+(P_0/c)q_A}.
 \label{eq:IR-A}
\end{align}

For prescription $B$, the initial effective quartic coupling is $\leff(0)=\lambda_0/Z_0^2$. Integration of the metric beta functions gives
\begin{equation}
 \xe^B-c=-\frac{c}{Z_0}q_B,\qquad
 \xp^B=c\left(\frac{Z_0}{q_B}-1\right).
 \label{eq:IR-B}
\end{equation}
Thus $\xe^{A,B}\to c$ as $t\to-\infty$, while $(\xp^A,\xg^A)\to(0,c)$ and $\xp^B\to+\infty$. The canonical quartics are $\lambda_{\mathrm{eff}}^A=\lambda_0q_A^3/Z_A^2$ and $\lambda_{\mathrm{eff}}^B=\lambda_0q_B^3/Z_0^2$, and both tend to zero.

For $-c<P_0<0$, both prescriptions initially have positive gravitational coupling. The reduced coefficients $f_A=-(\xp^A+\xg^A)\phi^2$ and $f_B=-\xp^B\phi^2$ vanish at $q_A=1/(1-P_0/c)$ and $q_B=Z_0$, respectively. These crossings occur at finite negative $t$ and make the Einstein-frame transformation singular. In prescription $B$, the crossing also has $\xp^B=0$: the original connection equation ceases to determine $\Gamma_\star$, and the auxiliary Hessian in Eq.~\eqref{eq:schur} is not invertible. In prescription $A$, $\xp^A$ remains nonzero at the zero of the total gravitational coefficient. The running formulas can be continued formally, but their common infrared endpoint lies beyond a single regular positive-gravity branch.

At the endpoint of prescription $A$, $\xp^A=0$, the connection decouples from the two-derivative sector. In prescription $B$, $\xp^B\to+\infty$ approaches the boundary of the coupling map. The corresponding compensator limit uses the field rescaling of Sec.~\ref{sec:reduction}. Since $Z_B=Z_0/q_B$, the original quartic is $\lambda_B=Z_B^2\lambda_{\mathrm{eff}}^B=\lambda_0q_B$. On the eventual $\xp^B>0$ branch,
\[
 \bar\lambda_B=\frac{\lambda_B}{(\xp^B)^2}
 =\frac{\lambda_0q_B^3}{c^2(Z_0-q_B)^2}\longrightarrow0.
\]
The trajectory therefore reaches the zero-potential compensator limit, rather than the fixed nonzero-$\bar\lambda$ limit used to obtain Eq.~\eqref{eq:R2-endpoint}.

\section{Summary and conclusions}
\label{ref:summary}

We have studied the relation between conformal coupling and scalar dynamics in both the metric formulation and the Palatini formulation. By considering a single scale-invariant scalar with a $\phi^2  {\cal R}$ coupling and a canonical kinetic term, we demonstrated that the Palatini trace of the energy-momentum tensor has no explicit dependence on $\xp$ when evaluated on the scalar equation of motion, and assumes the same form as in the minimally coupled case. The curvature term is Weyl invariant in isolation for an independent connection and as such cannot cancel the variation of the kinetic term.

Eliminating the connection produces an equivalent metric theory with kinetic coefficient $Z=1+\xp/\xc$. For $Z>0$, canonical normalization gives the effective curvature coupling $\xe=\xp/Z$; the other branches require separate treatment. The pole at $\xp=-\xc$ does not correspond to a loss of the scalar degree of freedom, because the full scalar--metric kinetic combination remains $\mathcal D_{\rm red}=k=1$. For $\xp<0$, the Einstein-frame theory contains one minimally coupled massless scalar and a cosmological constant. Within the ungauged model considered here, a Weyl compensator is obtained only after a field rescaling and an infinite-coupling limit.

Going beyond the tree level, we computed the scalar counterterms on a fixed metric-affine background. The generated $\phi^2R[g]$ coupling $\xg$ has the usual conformal shift in its beta function, while the independent-curvature coupling $\xp$ runs homogeneously. We also identified the required curvature-squared counterterms. Eliminating the auxiliary connection before quantizing includes its dependence on the scalar in the fluctuation operator. The resulting scalar--connection mixing accounts for the different Hessians and beta functions in the two prescriptions. For positive perturbative quartics, both admit the same formal one-loop infrared limit $(\xe,\leff)\to(1/6,0)$, reached through different limits of the original couplings. Within the two-derivative gravitational description, trajectories starting with positive gravitational coupling encounter $f=0$ at a finite scale before reaching this endpoint. The reduced trajectory approaches a zero-potential compensator limit, and approach to the free metric conformal coupling does not remove the gravitational trace anomaly.

Our calculations do not include quantum metric fluctuations. Their inclusion requires a gravitational effective action with higher-curvature terms. Such an analysis would determine the gravitational contributions to the curvature-coupling beta functions, test whether the infrared behavior persists, and allow the region near $f=0$ to be examined beyond the two-derivative approximation.

\section*{Acknowledgments}
This work was supported by the Estonian Research Council via the Center of Excellence TK202, ``Foundations of the Universe''. A.~K. was also supported by the Estonian Research Council grants TARISTU24-TK10 and TARISTU24-TK3. S.~K. was supported by the Estonian Research Council Mobilitas~3.0 incoming postdoctoral grant MOB3JD1233, ``Inflationary Nonminimal Models: An Investigative Exploration''.

\appendix
\section{Connection elimination and the energy-momentum trace}
\label{app:connection}
In this appendix, we demonstrate how the connection can be classically eliminated. We will also verify that the reduced theory correctly reproduces the original energy-momentum trace.

We begin by setting $H^{\mu\nu}=\sqrt{-g}\,f g^{\mu\nu}$. For a symmetric connection, it holds that
\begin{equation}
 \delta R_{\mu\nu}
 =\nabla^\Gamma_\lambda\delta\Gamma^\lambda{}_{\mu\nu}
 -\nabla^\Gamma_\nu\delta\Gamma^\lambda{}_{\mu\lambda}.
 \label{eq:delta-R}
\end{equation}
After the usual variation, the equation for $H^{\mu\nu}$ is
\begin{equation}
 -\nabla^\Gamma_\lambda H^{\mu\nu}
 +\delta_\lambda^{(\nu}\nabla^\Gamma_\rho H^{\mu)\rho}=0.
 \label{eq:connection-full}
\end{equation}
Contracting gives us $(D-1)\nabla^\Gamma_\rho H^{\mu\rho}/2=0$, and hence Eq.~\eqref{eq:compat}. 

On a fixed-sign $f\ne0$ patch, $H^{\mu\nu}$ is a multiple of $\sqrt{-h}h^{\mu\nu}$, with $h$ given by Eq.~\eqref{eq:hmetric}. For $D>2$, the torsion-free solution for the connection becomes
\begin{equation}
 \Gamma^\lambda{}_{\mu\nu}
 =\{^{\lambda}_{\mu\nu}\}_g
 +\frac1{D-2}\left(
 \delta^\lambda_\mu\partial_\nu\ln|f|
 +\delta^\lambda_\nu\partial_\mu\ln|f|
 -g_{\mu\nu}\partial^\lambda\ln|f|\right).
 \label{eq:connection-solution}
\end{equation}
This equation can be used to relate the generalized curvature scalar with the metric curvature scalar as given in Eq.~\eqref{eq:R-reduced}. Further integrating by parts gives
\begin{equation}
 \int\dd^Dx\sqrt{-g}\, \frac{f}{2}\Rc
 =\int\dd^Dx\sqrt{-g}\left[
 \frac{f}{2} R[g]+\frac{D-1}{2(D-2)}\frac{(\partial f)^2}{f}\right],
 \label{eq:IBP}
\end{equation}
up to the specified boundary term.

We can further check these results by taking the energy-momentum trace. An action in the metric formulation  with constant kinetic coefficient $Z$, curvature term $-\xp R\phi^2/2$, and arbitrary $V$ has scalar equation $Z\Box_g\phi-\xp R\phi-V'=0$. The energy-momentum trace is 
\begin{align}
 T_{\mathrm{red}}
 &=-\Delta Z(\partial\phi)^2-\Delta\xp R\phi^2
 +(D-1)\xp\Box_g\phi^2-DV \nonumber\\
 &=\left[(D-1)\xp-\frac\Delta2 Z\right]\Box_g\phi^2
 +\Delta\phi V'-DV.
 \label{eq:tracecheck}
\end{align}
Taking $Z=1+\xp/\xc$, we find that the first bracket equals $-\Delta/2$, which reproduces the original Palatini trace, something which also remains valid at $Z=0$.

\section{Reduced scalar Hessian and beta functions}
\label{app:hessian}

We present the scalar-loop calculation after eliminating the connection in more detail. First, let $A$ denote the independent components of the torsion-free connection, and define $A_\star[\phi]$ by $S_{,A}=0$ at fixed metric. By functionally differentiating, we find
\begin{equation}
 \frac{\delta A_\star}{\delta\phi}
 =-(S_{,AA})^{-1}S_{,A\phi},\qquad
 \frac{\delta^2 S[\phi,A_\star[\phi]]}{\delta\phi^2}
 =S_{,\phi\phi}-S_{,\phi A}(S_{,AA})^{-1}S_{,A\phi}.
 \label{eq:schur-proof}
\end{equation}
After integration by parts, we find that the connection enters the original action only algebraically. The connection Hessian $S_{,AA}$ can therefore be inverted separately at each spacetime point on a regular patch with $f\ne0$.

Alternatively, one can obtain the Schur complement of $S_{,AA}$ by taking the second variation of the action after eliminating the connection. For $D=4$, the action is
\begin{equation}
 S_{E,\mathrm{red}}=\int\dd^4x\sqrt g\left[
 \frac12 Z(\partial\phi)^2+\frac12\xp R\phi^2
 +\frac\lambda4\phi^4\right],\qquad Z=1+6\xp.
 \label{eq:Euclidean-reduced}
\end{equation}
Decomposing $\phi=\bar\phi+\eta$, we find the quadratic part to be
\begin{equation}
 S^{(2)}_{E,\mathrm{red}}
 =\frac12\int\dd^4x\sqrt g\,
 \eta\left[-Z\Box_g+\xp R+3\lambda\bar\phi^2\right]\eta.
 \label{eq:quadratic-reduced}
\end{equation}
If we identify $\eta_c=\sqrt Z\,\eta$ and $\bar\varphi=\sqrt Z\,\bar\phi$, then Eq.~\eqref{eq:reduced-operator} follows directly. After canonical normalization, the reduced scalar fluctuation operator matches the standard metric scalar operator with curvature coupling $\xe$ and quartic coupling $\leff$. In this calculation we have implicitly included the dependence of $\Gamma_\star[g,\phi]$ on $\phi$, which supplies the mixed
scalar--connection contribution in Eq.~\eqref{eq:schur-proof}.

For completeness, we express the reduced theory's beta functions in terms of the original couplings $(\xp,\lambda)$ using the inverse coupling map. Writing $c=1/6$, this map on the branch $Z>0$ is
\begin{equation}
 \xp=\frac{c\xe}{c-\xe},\qquad
 Z=\frac{c}{c-\xe},\qquad \lambda=Z^2\leff.
 \label{eq:inverse-map}
\end{equation}
The standard metric beta functions are
\begin{equation}
 \beta_{\xe}=\frac{3\leff}{8\pi^2}(\xe-c),\qquad
 \beta_{\leff}=\frac{9\leff^2}{8\pi^2}.
 \label{eq:metric-betas-app}
\end{equation}
Applying the chain rule gives
\begin{equation}
 \beta^{\mathrm{red}}_{\xp}=Z^2\beta_{\xe}
 =-\frac{3\leff}{8\pi^2}(\xp+c),\qquad
 \frac{\dd Z}{\dd t}=-\frac{3\leff}{8\pi^2}Z.
 \label{eq:pullback-xi}
\end{equation}
The running of $\lambda=Z^2\leff$ includes contributions from both
$Z$ and $\leff$:
\begin{equation}
 \beta^{\mathrm{red}}_\lambda
 =2Z\leff\frac{\dd Z}{\dd t}+Z^2\beta_{\leff}
 =\frac{3Z^2\leff^2}{8\pi^2}
 =\frac{3\lambda^2}{8\pi^2Z^2}.
 \label{eq:pullback-lambda}
\end{equation}

\bibliographystyle{utphys_fixed}
\bibliography{refs}

\end{document}